\documentclass[journal]{IEEEtranTIE}
\usepackage{graphicx}
\usepackage{cite}
\usepackage{picinpar}
\usepackage{amsmath,amssymb,amsfonts}
\usepackage{algorithmic}
\usepackage{url}
\usepackage[latin1]{inputenc}
\usepackage{colortbl}
\usepackage{soul}
\usepackage{multirow}
\usepackage{pifont}
\usepackage{color}
\usepackage{alltt}
\usepackage[hidelinks]{hyperref}
\usepackage{enumerate}
\usepackage{siunitx}
\usepackage{breakurl}
\usepackage{epstopdf}
\usepackage{pbox}
\usepackage{mathptmx}
\usepackage{bm}
\allowdisplaybreaks[4]
\graphicspath{{./figures/}}
\DeclareMathAlphabet{\pazocal}{OMS}{zplm}{m}{n}
\begin{document}
\title{On Power Control of Grid-Forming Converters: Modeling, Controllability, and Full-State Feedback Design}

\author{
	\vskip 1em
	
	Meng Chen, \emph{Member, IEEE},
	Dao Zhou, \emph{Senior Member, IEEE},
	Ali Tayyebi,
         Eduardo Prieto-Araujo, \emph{Senior Member, IEEE},
         Florian D$\ddot{\rm o}$rfler, \emph{Senior Member, IEEE},
	and Frede Blaabjerg, \emph{Fellow, IEEE}

	
		

}

\maketitle
	
\begin{abstract}
The popular single-input single-output control structures and classic design methods (e.g., root locus analysis) for the power control of grid-forming converters have limitations in applying to different line characteristics and providing favorable performance. This paper studies the grid-forming converter power loops from the perspective of multi-input multi-output systems. First, the error dynamics associated with power control loops (error-based state-space model) are derived while taking into account the natural dynamical coupling terms of the power converter models. Thereafter, the controllability Gramian of the grid-forming converter power loops is studied. Last, a full-state feedback control design using only the local measurements is applied. By this way, the eigenvalues of the system can be arbitrarily placed in the timescale of power loops based on predefined time-domain specifications. A step-by-step construction and design procedure of the power control of grid-forming converters is also given. The analysis and proposed method are verified by experimental results.
\end{abstract}


\markboth{IEEE TRANSACTIONS ON INDUSTRIAL ELECTRONICS}%
{}

\definecolor{limegreen}{rgb}{0.2, 0.8, 0.2}
\definecolor{forestgreen}{rgb}{0.13, 0.55, 0.13}
\definecolor{greenhtml}{rgb}{0.0, 0.5, 0.0}

\section{Introduction}

\IEEEPARstart{G}{rid-forming} converters are becoming vital to the power systems due to their ability to help stabilize the frequency and voltage. The control architecture of grid-forming converters is typically nested with multiple loops, e.g., inner cascaded voltage and current loops as well as the outer power loops. To simplify the analysis and design, the cascaded loops are usually designed with higher bandwidths than those of the power loops. As a result, the fast dynamics of the cascaded loops can be neglected when focusing on the design of the power loops \cite{Rosso2021}. Until now, several grid-forming power controls have been proposed, e.g., droop control\cite{Song2021,Mohammed2022}, virtual synchronous generator (VSG) control \cite{Chen2021b}, virtual oscillator control\cite{Ali2020}, matching control \cite{Tayyebi2020}, power synchronization control \cite{Quan2020a,Zhang2021}, hybrid angle control\cite{tayyebi2022hybrid}, etc., as well as their various improved forms \cite{Chen2022a}. Nevertheless, there still are many concerns which are worth studying in detail.

First, the control performance is limited by the line characteristics. Usually, an inductive or resistive line is assumed in the aforementioned controls, where, as a result, the frequency and voltage can be controlled by two decoupled single-input single-output (SISO) loops, e.g., $p$-$f$ and $q$-$V$ loops for an inductive line, respectively\cite{Zhong2013,Deng2021,Deng2022}. However, this decoupled control architecture will inevitably exert additional restraints on both the system and controller parameters \cite{Wu2016}. Further, the decoupling is just an approximate rather than exact result. More important, in a complex line, the frequency, voltage, active and reactive powers are tightly coupled with each other, which may lead to large errors and suboptimal dynamics performance when using two SISO control loops. In the literature, two methods have been used to solve this problem. Most usually, a virtual impedance can be added to enlarge the equivalent inductance \cite{Liu2020}. Therefore, the $p$-$f$ and $q$-$V$ relationships can still be used for a complex line. Nevertheless, the virtual impedance should be designed carefully to obtain favorable stability, performance, and avoid undesired wind-up behavior \cite{Ahmed2022}. Meanwhile, this method is still relies on an approximate decoupling. The other method is to use multi-input multi-output (MIMO) control structures \cite{Rafiee2021,Chen2022}, e.g., the MIMO-GFM converter, where no assumption of decoupling needs to be made. Therefore, they are potentially effective for different line characteristics, although no detailed study has been performed until now.

As a second open point, the conventional parameter tuning is cumbersome, and the region of the achievable performance is narrow. Until now, classic design methods, e.g, root locus \cite{Chen2021} and frequency-domain analysis\cite{Chen2019,Ambia2021}, have been widely used in the design and tuning of the grid-forming power control loops. On the one hand, these methods are manual and trial-and-error due to the fact that only one parameter can be tuned at a time and the influences of different parameters can be conflicting, which renders the control tuning a daunting task. On the other hand, the achievable performance is limited, e.g., the eigenvalues cannot be placed to the positions beyond the root loci. As a result, the required stability and time-domain performance, e.g., settling time ($T_{s}$), percentage overshoot ($P.O.$), etc., may not be guaranteed. Some other design methods such as $\pazocal{H}_{\infty}$ synthesis have also been used, which are convenient to tune a multi-parameters system and to achieve an optimal performance \cite{Huang2020}. Their performance depends on the deployed weighting functions which are hard to tune from time-domain specifications.

Third and finally, some basic properties have not been studied, e.g., the controllability \cite{Ogata2009}. As a controlled system, the controllability determines whether there exist suitable inputs that can transfer the states of the grid-forming power loops from any initial values to the the equilibrium. The controllability is of great importance in control design especially for state-space models, which can be used to explain why a specific control structure is effective. Although the existing methods do not mention it, some of them do depend on controllability, i.e., the states can be controlled by the inputs.  

Motivated by the aforementioned analysis, a full-state feedback based-power control has been proposed for grid-forming converters in \cite{chen2022power}, where only the basic idea and results have been given. In this paper, the details will be discussed from the perspectives of modeling, controllability, and design. Experimental validations with different line characteristics will also be presented. In summary, the proposed method has the following advantages and contributions.
\begin{enumerate}
	\item The established error-based state-space model is a MIMO system considering the natural coupling between the active and reactive power loops. Therefore, the analysis and controller design are expected to have favorable robustness to mixed line characteristics (i.e., resistive, inductive, and complex) and grid strength (i.e., small and large short-circuit ratio (SCR)).
	\item The controllability Gramian is studied in detail for the first time, which provides a theoretical basis to the subsequent controller design.
       \item A full-state feedback control design with only local measurements is proposed, where the eigenvalues can be placed at any position within the timescale of the power loops. A step-by-step parameters design procedure based on the predefined time-domain performance is provided.
\end{enumerate}
 
The remainder of paper is organized as follows. The error-based model of the grid-forming converter power loops is built in Section II. In Section III, an analysis on the controllability is carried out with the details of the full-state feedback-based power control being given in Section IV. In Section V, experimental results are shown, and finally, conclusions are given in Section VI.

\section{Error-Based State-Space Modeling of Grid-Forming Converters Power Loops}

\begin{figure}[!t]\centering
	\includegraphics[width=\columnwidth]{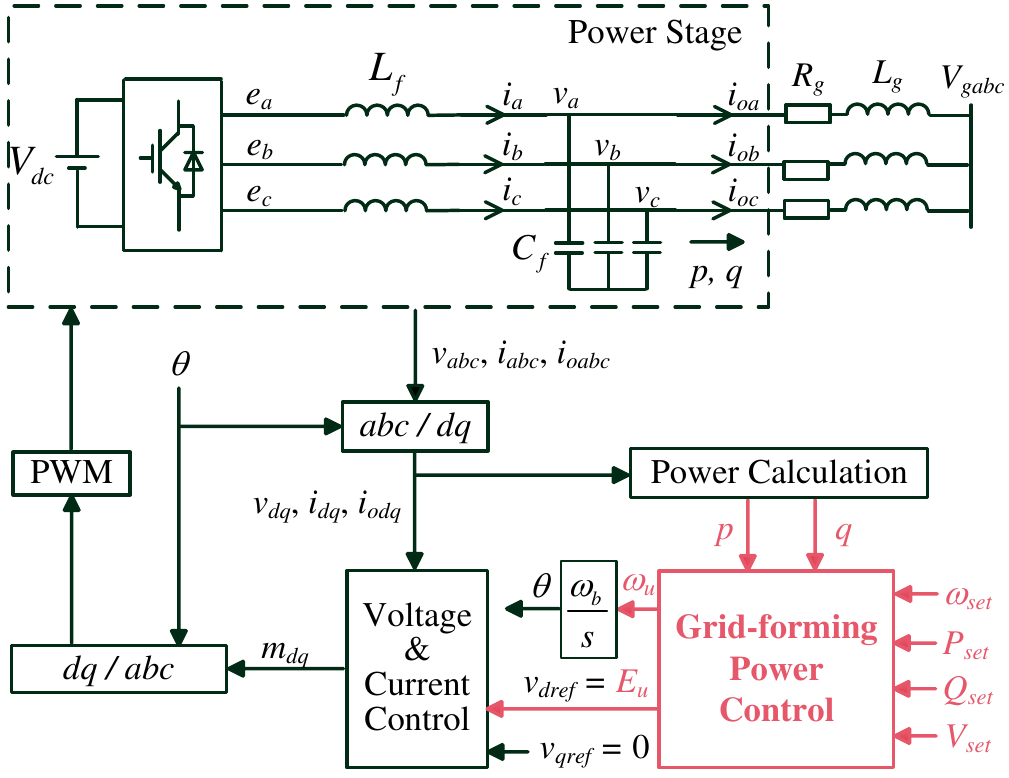}
	\caption{General configuration of grid-forming converters with outer power control and inner cascaded control.}\label{VSC}
\vspace{-10pt}
\end{figure}

Fig. \ref{VSC} shows the general configuration of a grid-forming converter, where the power stage consists of a three-phase converter and an LC filter. $L_f$ and $C_f$ are the filter inductor and capacitor, respectively. $L_g$ and $R_g$ are the inductor and resistor of the line to the power grid. The grid-forming power control is responsible for providing the frequency and voltage references, i.e., $\omega_u$ and $E_u$, based on the calculated output active and reactive powers of the converter, i.e., $p$ and $q$. Thereafter, a typical cascaded voltage and current loop is used to generate the modulation signals by regulating the capacitor voltages and inductor currents of the filter.

When considering a general line impedance ($Z_g=R_g+jX_g\neq 0$), which may be complex, the output powers $p$ and $q$ may couple with each other by the following equations: \cite{Mohammed2022}. 
\begin{align}
\label{p}
&p=\frac{V^2R_g+VV_g(X_gsin\delta-R_gcos\delta)}{R_g^2+X_g^2}\\
&q=\frac{V^2X_g-VV_g(R_gsin\delta+X_gcos\delta)}{R_g^2+X_g^2}
\end{align}

Here $V$ and $V_g$ are the voltage magnitudes of the capacitor and grid, respectively. Moreover, $\delta$ is the angle separation between the vectors of capacitor and grid voltage defined as
\begin{align}
\label{delta}
\dot\delta=\omega_b\omega-\omega_b\omega_g,
\end{align}
where $\omega$ and $\omega_g$ are the voltage frequencies of the capacitor and grid, and $\omega_b$ is the base value of the frequency. 

The small-signal model of (\ref{p})-(\ref{delta}) can be derived as
\begin{align}
\label{delta_p}
&\Delta p=k_{p\delta}\Delta\delta+k_{pV}\Delta V\\
\label{delta_q}
&\Delta q=k_{q\delta}\Delta\delta+k_{qV}\Delta V\\
\label{ddelta_delta}
&\Delta\dot\delta=\omega_b\Delta\omega-\omega_b\Delta\omega_g
\end{align}
where
\begin{align}
\label{kpdelta}
&k_{p\delta}=\left.\frac{\partial p}{\partial\delta}\right|_{\delta_0,V_0}=\frac{V_0V_g(R_gsin\delta_0+X_gcos\delta_0)}{R_g^2+X_g^2}\\
&k_{pV}=\left.\frac{\partial p}{\partial V}\right|_{\delta_0,V_0}=\frac{2V_0R_g+V_g(X_gsin\delta_0-R_gcos\delta_0)}{R_g^2+X_g^2}\\
&k_{q\delta}=\left.\frac{\partial q}{\partial\delta}\right|_{\delta_0,V_0}=\frac{V_0V_g(X_gsin\delta_0-R_gcos\delta_0)}{R_g^2+X_g^2}\\
\label{kqv}
&k_{qV}=\left.\frac{\partial q}{\partial V}\right|_{\delta_0,V_0}=\frac{2V_0X_g-V_g(R_gsin\delta_0+X_gcos\delta_0)}{R_g^2+X_g^2}
\end{align}
and the subscript "0" represent the variables corresponding to the used steady-state operation point to linearize the model.

Due to the much larger bandwidths of the cascaded loops, their quick dynamics can be neglected to obtain
\begin{align}
\label{cascaded}
\left[\begin{matrix}
	\Delta\omega&\Delta V
\end{matrix}\right]^T=\left[\begin{matrix}
	\Delta\omega_u&\Delta E_u
\end{matrix}\right]^T
\end{align}
where (\ref{delta_p})-(\ref{cascaded}) consist of the regular open-loop model of the grid-forming power loops. 

To share the power among multiple inverters in a potential islanded operation mode or to enable coordinated power-sharing among different generation units in a transmission grid, the steady-state droop characteristics are expected to be included in the grid-forming control. It is widely recognized that the following $p$-$f$ and $q$-$V$ droops should be used for an inductive line
\begin{align}
\label{p_f droop}
&\omega_u-\omega_{set}=D_p(P_{set}-p)\\
\label{q_v droop}
&V-V_{set}=D_q(Q_{set}-q),
\end{align}
where $D_p$ and $D_q$ are the droop coefficients, the subscript ``set" represents the variables corresponding to the set-point. On the contrary, the following $p$-$V$ and $q$-$f$ droops should be used for a resistive line:
\begin{align}
\label{p_V droop}
&V-V_{set}=D_p(P_{set}-p)\\
\label{q_f droop}
&\omega_u-\omega_{set}=D_q(Q_{set}-q)
\end{align}

As an example, in this paper, the $p$-$f$ and $q$-$V$ droops of (\ref{p_f droop}) and (\ref{q_v droop}) are used in the modeling. Nevertheless, the same method can be applied to a system with the $p$-$V$ and $q$-$f$ steady-state droops. In the following, it will be illustrated that the proposed control can be effective for a complex line as well. Therefore, we define the droop output $\bm y$ and its reference $\bm{y_{ref}}$ as
\begin{align}
\label{y}
&\bm y=\left[\begin{matrix}
	y_1&y_2
\end{matrix}\right]^T=\left[\begin{matrix}
	\omega_u+D_pp&V+D_qq
\end{matrix}\right]^T\\
&\bm{y_{ref}}=\left[\begin{matrix}
	y_{1ref}&y_{2ref}
\end{matrix}\right]^T=\left[\begin{matrix}
	\omega_{set}+D_pP_{set}&V_{set}+D_qQ_{set}
\end{matrix}\right]^T
\end{align}  

\begin{figure}[!t]\centering
	\includegraphics[width=\columnwidth]{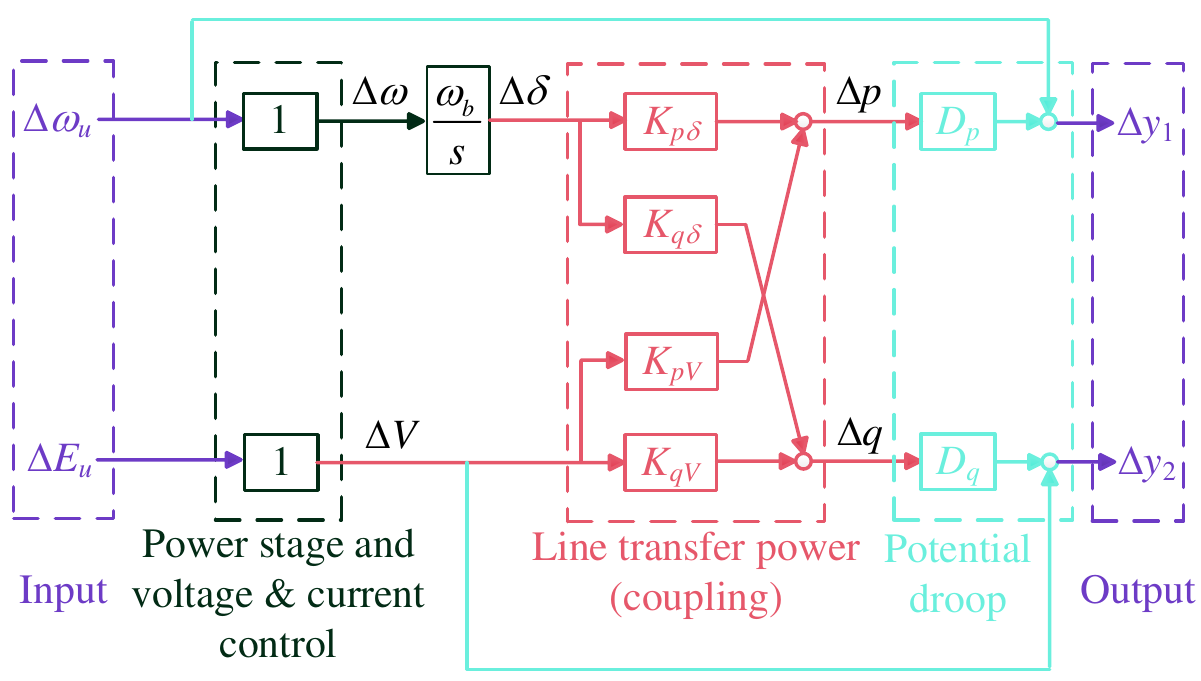}
	\caption{Open-loop small-signal model of grid-forming power loops with droop characteristics.}\label{extended_open}
\vspace{-10pt}
\end{figure}

Combining (\ref{delta_p})-(\ref{cascaded}) and (\ref{y}), the open-loop small-signal model of the grid-forming converter power loops with droop characteristics can be derived as
\begin{align}
\label{ddelta}
&\Delta\dot\delta=\left[\begin{matrix}
	\omega_b&0
\end{matrix}\right]\left[\begin{matrix}
	\Delta\omega_u& \Delta E_u
\end{matrix}\right]^T-\omega_b\Delta\omega_g\\
\label{output}
&\Delta\bm y=\left[\begin{matrix}
	D_pk_{p\delta}\\D_qk_{q\delta}
\end{matrix}\right]\Delta\delta+\left[\begin{matrix}
	1&D_pk_{pV}\\0&1+D_qk_{qV}
\end{matrix}\right]\left[\begin{matrix}
	\Delta\omega_u\\ \Delta E_u
\end{matrix}\right]
\end{align}
which is illustrated in Fig. \ref{extended_open}. Clearly, it represents a two-input two-output system. The grid-forming control aims to find proper closed-loop controllers so that the outputs $\bm y$ can track the reference $\bm{y_{ref}}$ with zero steady-state errors by favorable dynamics. Therefore, in the following, we establish the error-based model and transfer the tracking problem to a regulator problem for the convenience of analysis and design.

To this end, define the error signals as
\begin{align}
\Delta\bm e=\Delta\bm y-\Delta\bm{y_{ref}},
\end{align}
where, for the step references, there is
\begin{align}
\label{de}
\Delta\dot{\bm e}=\Delta\dot{\bm y}.
\end{align}

Placing (\ref{output}) into (\ref{de}) yields
\begin{align}
\label{de_new1}
\Delta\dot{\bm e}=\left[\begin{matrix}
	D_pk_{p\delta}\\D_qk_{q\delta}
\end{matrix}\right]\Delta\dot\delta+\left[\begin{matrix}
	1&D_pk_{pV}\\0&1+D_qk_{qV}
\end{matrix}\right]\left[\begin{matrix}
	\Delta\dot\omega_u\\ \Delta\dot E_u
\end{matrix}\right].
\end{align}

Define the following intermediate state $\Delta z$ and input $\Delta\bm u$
\begin{align}
\label{intermediate}
\Delta z=\Delta\dot\delta,~\Delta\bm u=\left[\begin{matrix}
\Delta u_1&\Delta u_2
\end{matrix}\right]^T=\left[\begin{matrix}
\Delta\dot\omega_u&\Delta\dot E_u
\end{matrix}\right]^T.
\end{align}
The error-based state-space model of the grid-forming converter power loops can then be stated as
\begin{align}
\label{error_model}
\Delta\dot{\bm x}=\bm A\Delta\bm x+\bm B\Delta\bm u
\end{align}
and the state vector $\Delta\bm x$, state matrix $\bm A$, and control matrix $\bm B$ are defined as
\begin{align}
\label{error_model_matrix}
\Delta\bm x=\left[\begin{matrix}
	\Delta\bm e\\\Delta z
\end{matrix}\right],~
\bm A=\left[\begin{matrix}
0&0&D_pk_{p\delta}\\0&0&D_qk_{q\delta}\\0&0&0
\end{matrix}\right],~\bm B=\left[\begin{matrix}
1&D_pk_{pV}\\0&1+D_qk_{qV}\\\omega_b&0
\end{matrix}\right]
\end{align}
\textbf{Remark~1}. \textit{Compared with the original model of (\ref{delta_p})-(\ref{cascaded}), the model of (\ref{error_model}) takes the droop error $\Delta\bm e$ and the frequency error $\Delta z$ as the states. As a result, once a closed-loop controller is designed to provide a favorable (stable) dynamic performance, the steady-state droop characteristics and the frequency synchronization will be asymptotically guaranteed (in presence of step disturbances) according to the internal model principle \cite{Dorf2010}.}

\section{Analysis on Controllability Gramian}

Before going on with the controller design, this section investigates controllability, i.e., whether the states can be controlled by the inputs. The controllability governs the existence of the studied control design problem in theory. For a given state-space model, if it is not controllable, a proper input may never be found whatever static or dynamic gains are chosen. The analysis in this section uses the following lemma.\\
\textbf{Lemma~1} \cite{Chen1999}. \textit{Given the state-space model of (\ref{error_model}), where \textbf{A} is an n$\times$n matrix, the system is controllable if and only if the controllability Gramian $\bm {P_T}$ is nonsingular, where}
\begin{align}
\label{pt}
\bm{P_T}\triangleq\int_0^te^{\bm A\tau}\bm B\bm B^Te^{\bm A^T\tau}d\tau.
\end{align}

It is noticed, according to (\ref{error_model_matrix}), that
\begin{align}
\bm A^2=0,
\end{align}
which yields that
\begin{align}
e^{\bm A\tau}=\bm I+\bm A\tau,~e^{\bm A^T\tau}=\bm I+\bm A^T\tau
\end{align}

Therefore, the controllability Gramian (\ref{pt}) is equivalent to
\begin{align}
\bm{P_T}\triangleq\int_0^t(\bm I+\bm A\tau)\bm B\bm B^T(\bm I+\bm A^T\tau)d\tau,
\end{align}
where the final result is shown as (\ref{pt_final}) on the top of next page. To guarantee the nonsingularity of $\bm{P_T}$, its determinant should be nonzero, i.e.,
\begin{figure*}
\begin{footnotesize}
\begin{align}
\label{pt_final}
\bm{P_T}=\left[\begin{matrix}
1/3\omega_b^2D_p^2k_{p\delta}^2t^3+\omega_bD_pk_{p\delta}t^2+(1+D_p^2k_{pV}^2)t & 1/3\omega_b^2D_pD_qk_{p\delta}k_{q\delta}t^3+1/2\omega_bD_qk_{q\delta}t^2+D_pk_{pV}(1+D_qk_{qV}^2)t & 1/2\omega_b^2D_pk_{p\delta}t^2+\omega_bt\\
1/3\omega_b^2D_pD_qk_{p\delta}k_{q\delta}t^3+1/2\omega_bD_qk_{q\delta}t^2+D_pk_{pV}(1+D_qk_{qV}^2)t &1/3\omega_b^2D_q^2k_{q\delta}^2t^3+(1+D_qk_{qV})^2t & 1/2\omega_b^2D_qk_{q\delta}t^2\\
1/2\omega_b^2D_pk_{p\delta}t^2+\omega_bt& 1/2\omega_b^2D_qk_{q\delta}t^2&\omega_b^2t
\end{matrix}\right]
\end{align}
\end{footnotesize}
\vspace{-5ex}
\end{figure*}
\begin{align}
\label{controllability_intermediate1}
1/12D_p^2\omega_b^4(k_{p\delta}+D_qk_{p\delta}k_{qV}-D_qk_{pV}k_{q\delta})^2t^5\neq 0
\end{align}
\textbf{Remark~2}. \textit{The controllability Gramian is a quantitative metric. The further its determinant is bounded away from zero the easier it is to control the system. The controllability Gramian reflects how much energy is needed to control the system and can be used to construct a so-called minimal energy control, which will, however, not be further discussed in this paper.}
 
Substituting (\ref{kpdelta})-(\ref{kqv}) into (\ref{controllability_intermediate1}) yields
\begin{align}
\label{controllability_condition}
F_c\triangleq \frac{D_pV_0V_g(R_gsin\delta_0+X_gcos\delta_0-D_qV_g+2V_0D_qcos\delta_0)}{R_g^2+X_g^2}\neq 0
\end{align}
\textbf{Remark~3}. \textit{The condition (\ref{controllability_condition}) holds in usual as $D_pV_0V_g\neq 0$ in a normal operation and $2V_0cos\delta_0\ge V_g$ for a small $\delta_0$, which leads to $F_c>0$. Even through $F_c=0$ for some large $\delta_0$, when the system is deviated from the equilibrium ($\delta_0$, $V_0$) due to disturbances (both external disturbances and set-points changing), there is $F_c\neq 0$ and the system is controllable.}

Now consider a resistive line by setting $X_g=0$, where $F_c$ becomes
\begin{align}
F_c=\frac{D_pV_0V_g(R_gsin\delta_0-D_qV_g+2V_0D_qcos\delta_0)}{R_g^2}
\end{align}
which means that although the $p-f$ and $q-V$ droops of (\ref{p_f droop}) and (\ref{q_v droop}) are applied to a resistive line, the system may still be controllable. That is to say, the steady-state droop characteristics can be decoupled with the line characteristics. Similarly, it can be proved that the $p-V$ and $q-f$ droops of (\ref{p_V droop}) and (\ref{q_f droop}) can also be applied to an inductive network. Nevertheless, these conclusions are only related to an ideal model (without physical restraints). It still prefers to use the $p-V$ and $q-f$ droops to a resistive line. Otherwise, the grid-forming converter may need to absorb large reactive power to output the required active power.\\
\textbf{Remark~4}. \textit{The condition (\ref{controllability_condition}) is also effective for different (non-zero) SCRs, i.e, the system can be controllable when connected into power grids with different strengths.}

\section{Full-State Feedback Design}
This section first presents how to construct the feedback loops based on the full-state feedback control when the controllability is clear. Then a step-by-step gains design based on the predefined time-domain specifications is given.

\subsection{Control Law Construction}
To construct the feedback laws, the following lemma is used.\\
\textbf{Lemma~2} \cite{Ogata2009}. \textit{Given the state-space model of (\ref{error_model}), the close-loop poles can be placed at arbitrary locations by the following control law if and only if the system is controllable.}
\begin{align}
\label{full_state_feedback}
\Delta\bm u=-\bm{K}\Delta\bm x
\end{align}
where $\bm K$ is a static gain matrix.

The studied error-based space-space model in this paper represents a system with three states and two inputs, according to \textbf{Lemma 2}, the gain matrix $\bm K$ can be expressed as
\begin{align}
\label{k}
\bm K=\left[\begin{matrix}
	k_{11}&k_{12}&k_{13}\\
	k_{21}&k_{22}&k_{23}
\end{matrix}\right]
\end{align}
where $k_{ij}$ is the control gain. Placing (\ref{k}) and the specific forms, corresponding to the grid-forming power loops, of $\Delta\bm u$ (\ref{intermediate}) and $\Delta\bm x$ (\ref{error_model_matrix}) into (\ref{full_state_feedback}) yields
\begin{align}
\left[\begin{matrix}
	\Delta\dot\omega_u\\\Delta\dot E_u
\end{matrix}\right]=\left[\begin{matrix}
k_{11}&k_{12}\\k_{21}&k_{22}
\end{matrix}\right]\left[\begin{matrix}
-\Delta e_1\\-\Delta e_2
\end{matrix}\right]-\left[\begin{matrix}
k_{13}\\k_{23}
\end{matrix}\right]\Delta z
\end{align}
where the actual inputs provided by the grid-forming power control, combing with the definition of $\Delta z$ (\ref{intermediate}), are derived as
\begin{align}
\left[\begin{matrix}
	\Delta\omega_u\\\Delta E_u
\end{matrix}\right]=-\int_0^t\left(\left[\begin{matrix}
k_{11}&k_{12}\\k_{21}&k_{22}
\end{matrix}\right]\left[\begin{matrix}
e_1\\e_2
\end{matrix}\right]\right)d\tau-\left[\begin{matrix}
k_{13}\\k_{23}
\end{matrix}\right]\Delta\delta
\end{align}

Until now, the design of the close-loop structure for the grid-forming power loops assumes that all the necessary signals are measurable. Nevertheless, $\Delta\delta$ may not be locally available due to the fact that $\Delta\omega_g$ is a remote disturbance signal. To solve this problem, the actual $\Delta\delta$ can be estimated using (\ref{delta_p}) and (\ref{delta_q}) as
\begin{align}
\Delta\hat\delta=k_p\Delta p-k_q\Delta q
\end{align}
where
\begin{align}
\label{kp_kq}
k_p=\frac{K_{qV}}{K_{p\delta}K_{qV}-K_{pV}K_{q\delta}},~k_q=\frac{K_{pV}}{K_{p\delta}K_{qV}-K_{pV}K_{q\delta}}
\end{align}

Finally, the complete close-loop small-signal model of the proposed full-state feedback-based grid-forming converter can be shown as Fig. 3.\\
\textbf{Remark~5}. \textit{By setting $k_{12}=k_{21}=k_{13}=k_{23}=0$, the proposed MIMO full-state feedback-based grid-forming structure becomes the traditional VSG assuming a ($p$, $f$) and ($q$, $V$) decoupling, i.e., the VSG control is a special case of the proposed structure with simplified SISO loops and partial-state feedback. However, these four parameters $k_{12}$, $k_{21}$, $k_{13}$, $k_{23}$ couple the active and reactive power loops. As a result, the proposed MIMO structure has good robustness to line impedance characteristics.}

\begin{figure*}[!t]\centering
	\includegraphics[width=\textwidth]{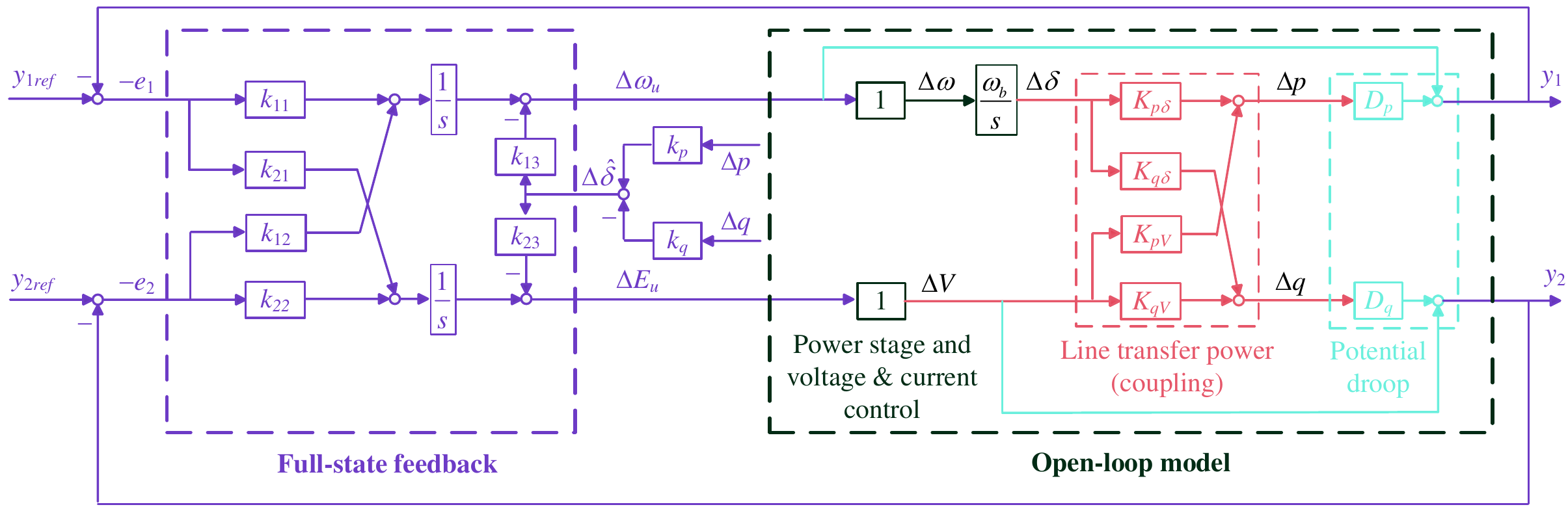}
	\caption{Close-loop small-signal model of proposed full-state feedback-based grid-forming converter.}\label{close_loop}
\vspace{-10pt}
\end{figure*}

\subsection{Parameters Design}

By applying the full-state feedback control law of (\ref{full_state_feedback}), the closed-loop state-space model of (\ref{error_model}) is derived as
\begin{align}
\Delta\dot{\bm x}=(\bm A-\bm{BK})\Delta\bm x
\end{align}
where its characteristic equation is
\begin{align}
\label{char}
	\left|\lambda\bm{I}-\bm A+\bm{BK}\right|=0
\end{align}

The characteristic equation (\ref{char}) has three eigenvalues, which can be placed at arbitrary locations according to \textbf{Lemma 2}. As a reasonable choice, we choose a pair of complex eigenvalue as the dominant ones and a real eigenvalue, which is far away from the dominant eigenvalues. Therefore, the characteristic equation should has the following form
\begin{align}
	(\lambda+a)(\lambda^2+2\xi\omega_n\lambda+\omega_n^2)=0
\end{align}
where $-a$ is a chosen real eigenvalue, $\xi$ and $\omega_n$ are the damping ratio and natural frequency of the chosen complex eigenvalues. Thereafter, the gain matrix $\bm K$ can be solved by
\begin{align}
\label{solve_k}
	\left|\lambda\bm{I}-\bm A+\bm{BK}\right|\equiv(\lambda+a)(\lambda^2+2\xi\omega_n\lambda+\omega_n^2)
\end{align}
which always has solutions due to \textbf{Lemma 2}. For a MIMO system, the solutions are usually not unique, where, we can use the degrees of freedom to provide good robustness \cite{Ogata2009}.

Moreover, time-domain performance is one of the commonly used indices in practice. For a dominant second-order system, i.e., assuming $-a$ is far in the left half plane, the time-domain performance has direct relationship with $\xi$ and $\omega_n$ \cite{Ogata2009}, e.g., by taking $P.O.$ and $T_s$ as examples in this paper,
\begin{align}
\label{overshoot}
&P.O.=e^{-(\xi/\sqrt{1-\xi^2})}\times100\%\\
\label{settle_time}
&T_s=\frac{4}{\xi\omega_n}
\end{align}
which implies that the dominant complex eigenvalues can be calculated by the predefined time domain performance.

According to the aforementioned discussion, a step-by-step parameter design procedure for the proposed full-state feedback-based grid-forming control can be summarized as follows.
\begin{itemize}
	\item step 1: Preparation. Before designing the parameters, it should know the predefined time-domain performance as defined by the expected locations of the eigenvalues ($\xi$, $\omega_n$, and $a$) according to (\ref{overshoot}) and (\ref{settle_time}).
	\item step 2: Linearization. Choosing the steady-state operation point ($\delta_0$, $V_0$) to get the linearized system parameters $K_{p\delta}$, $K_{pV}$, $K_{q\delta}$, $K_{qV}$, $\bm A$, and $\bm B$ according to (\ref{kpdelta})-(\ref{kqv}) and (\ref{error_model_matrix}).
	\item step 3: Controllability checking. The judgment is based on (\ref{controllability_condition}).
	\item step 4: Parameters calculation. $k_p$ and $k_q$ can be calculated according to (\ref{kp_kq}). $\bm K$ can be solved according to (\ref{solve_k}).
\end{itemize}
\textbf{Remark~6}. \textit{The proposed method is model-based, which relies on the parameters of the system. It does not narrow the application as the traditional popular methods of root locus analysis and frequency analysis rely on the same requirement. Moreover, as the proposed method can arbitrarily place the eigenvalues, a good stability margin can be guaranteed to improve the robustness on parameter variations. Even when some parameters are completely unknown, the proposed structure can potentially have better dynamics than the popular VSG control which is special case (referring to \textbf{Remark 5}).}\\    
\textbf{Remark~7}. \textit{\textbf{Lemma 2} has provided an exact condition to arbitrarily placed the eigenvalues of the system. Therefore, any additional loops to improve the stability are unnecessary. However, they may improve the transient performance.}\\
\textbf{Remark~8}. \textit{Although there is \textbf{Remark 7}, it is only related to the eigenvalues (poles) of the grid-forming power loops. Other issues may still need further study, e.g., issues related to the zeros, systems beyond Fig. \ref{VSC}, coupled power loops and cascaded loops, etc.}\\
\textbf{Remark~9}. \textit{For a controllable system, other methods rather than the used pole placement method may also be applied, e.g., the linear quadratic regulator (LRQ), to achieve an optimal design, which could be a future work.}

\section{Experimental Validation}

\begin{figure}[!t]\centering
	\includegraphics[width=\columnwidth]{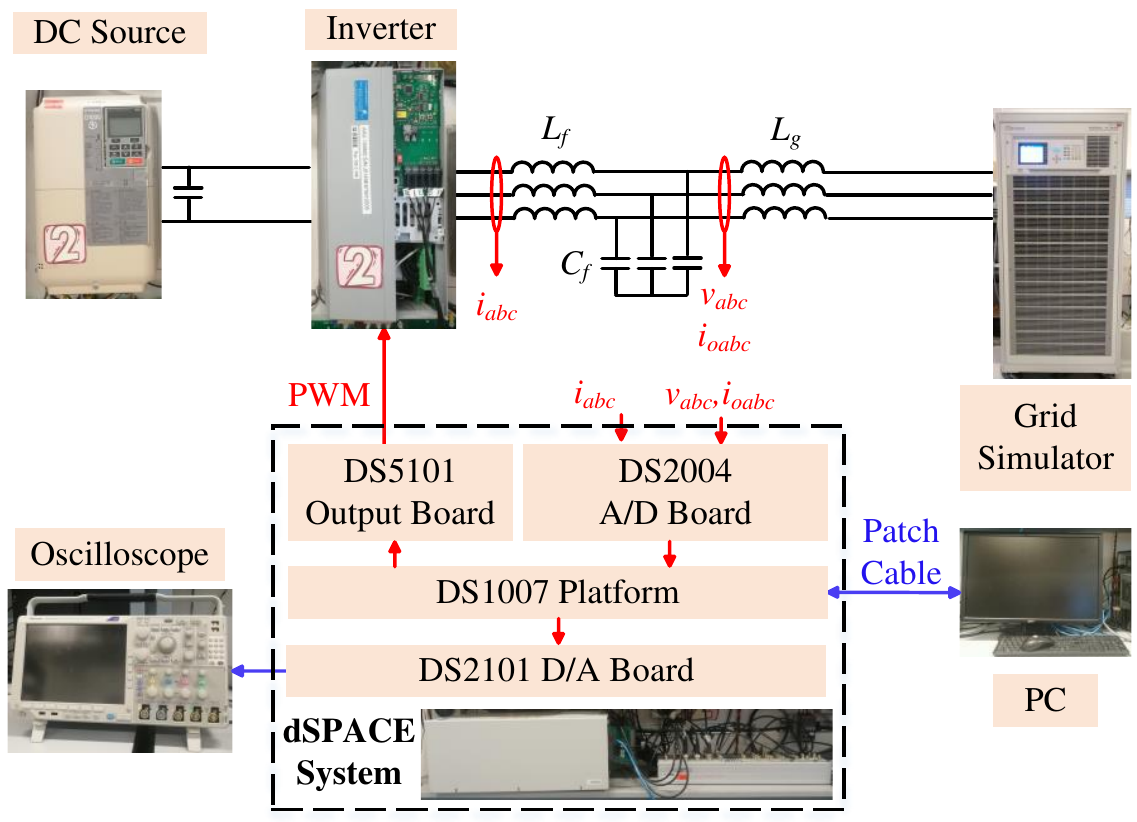}
	\caption{Experimental configuration of grid-forming converter.}\label{setup}
\vspace{-10pt}
\end{figure}

To verify the proposed full-state feedback control structure and parameter design for power loops of the grid-forming converter, this section will present some experimental results. The configuration of the setup is shown in Fig. \ref{setup}, where the power stage consists of a Danfoss drives system, an LCL filter and a Chroma 61845 grid simulator. The control is implemented by the DS1007 dSPACE system. Meanwhile, the DS2004 A/D board and DS2101 D/A board are used to collect the measurements and generate the output, respectively. The used key parameters are given in Table \ref{parameter} if there is no specific illustration. 

\begin{table}[!t]
	\renewcommand{\arraystretch}{1.3}
	\caption{Parameters of Experimental Setups}
	\centering
	\label{parameter}
	\resizebox{\columnwidth}{!}{
		\begin{tabular}{c l c}
			\hline\hline \\[-3mm]
			Symbol & Description & Value  \\ \hline
			$f_n$  & Nominal frequency & $100\pi$ rad/s \\
			$S_n $ & Nominal power &  5 kW  \\ 
			$V_n $ & Nominal line-to-line RMS voltage & 200 V \\
			$f_{sw}$ & Switching frequency & 10 kHz \\
			$\omega_g $ & Grid frequency & $100\pi$ rad/s (1 p.u.)  \\
			$V_g $ & Line-to-line RMS grid voltage & 200 V (1 p.u.) \\
			$L_g$  & Line inductor & 2.5 mH (0.0982 p.u.)\\
			$C_f$ & Filter capacitor & 15 $\mu$F (0.0377 p.u.)\\   
			$L_f$ & Filter inductor & 1.5 mH (0.0589 p.u.)\\
			$D_p$ & Droop coefficient of $P$-$f$ regulation & 0.01 p.u. \\ 
			$D_q$ &Droop coefficient of $Q$-$V$ regulation & 0.05 p.u.\\
			$\omega_{set}$&Frequency reference & 1 p.u. \\
			$P_{set}$ & Active power reference & 0.5 p.u.\\
			$Q_{set}$ & Reactive power reference & 0 p.u.\\
			$V_{set}$ & Voltage magnitude reference & 1 p.u.\\[1.4ex]
			\hline\hline
		\end{tabular}
	}
\end{table}

\subsection{Test on Different Predefined Time-domain Performance}
\begin{figure}[!t]\centering
	\includegraphics[width=\columnwidth]{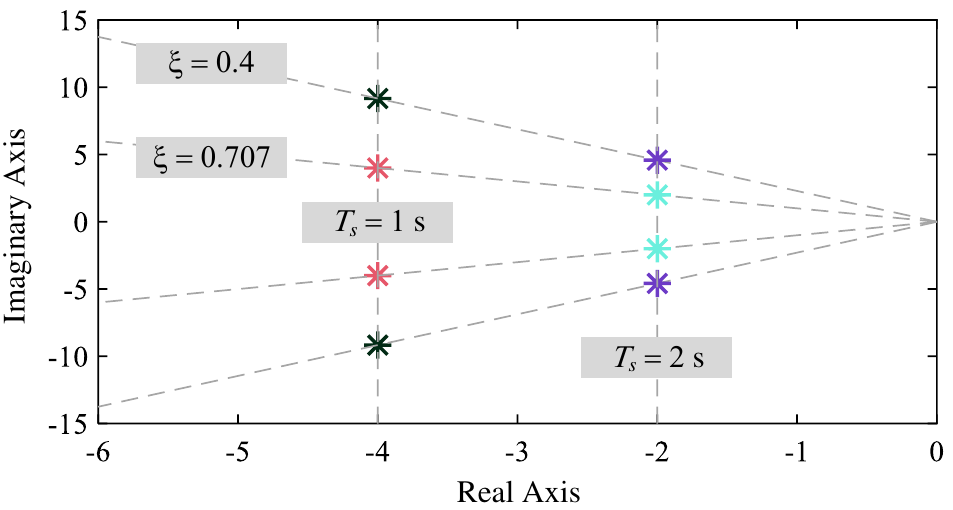}
	\caption{Chosen dominant complex eigenvalues.}\label{eig}
\vspace{-10pt}
\end{figure}

In this section, the dynamics of the proposed controller and parameter design method is tested with different predefined time-domain performance (i.e., $P.O.$ and $T_s$). The step-by-step parameter design is exerted as follows.
\begin{itemize}
	\item step 1: Preparation.\\
Four different studied cases are used as shown in Table \ref{case_time_domain}, where the third eigenvalue is always located far away from the dominant complex ones. The locations of the dominant complex eigenvalues are also presented in Fig. \ref{eig}.
	\item step 2: Linearization.\\
The steady-state operation point calculated based on Table \ref{parameter} is ($\delta_0$, $V_0$) = (0.0491, 0.9996). Thereafter, the linearized system parameters can be derived as follows.
\begin{align}
&(K_{p\delta},K_{pV},K_{q\delta},K_{qV}) = (10.1695,0.5002,0.5,10.1899)\\
&\bm A=\left[\begin{matrix}
			0&0&0.1017\\
			0&0&0.025\\
			0&0&0
		\end{matrix}\right],~\bm B=\left[\begin{matrix}
			1&0.005\\
			0&1.5095\\
			314.1593&0
		\end{matrix}\right]
\end{align}
	\item step 3: Controllability checking.\\
With the calculated ($\delta_0$, $V_0$), there is $F_c=0.1534>0$, which implies that the system is completely state controllable.
	\item step 4: Parameters calculation.\\
 The calculated control gains of different cases are listed in Table \ref{designed_parameter}.
\end{itemize}

\begin{table}[!t]
	\renewcommand{\arraystretch}{1.3}
	\caption{Studied Cases to Place Eigenvalues}
	\centering
	\label{case_time_domain}
	\resizebox{\columnwidth}{!}{
		\begin{tabular}{c l cc}
			\hline\hline \\[-3mm]
			Cases & Damping Ratio $\xi$ & Settling Time $T_s$  & Third eigenvalue $a$\\ \hline
			1  & 0.4 & 1 s & -20\\
			2 & 0.4 &  2 s & -20  \\ 
			3 & 0.707 & 1 s &-20 \\
			4 & 0.707 & 2 s & -20 \\[1.4ex]
			\hline\hline
		\end{tabular}
	}
\end{table}

\begin{table*}[!t]
	\renewcommand{\arraystretch}{1.3}
	\caption{Designed parameters of the corresponding cases}
	\centering
	\label{designed_parameter}
	\resizebox{0.7\textwidth}{!}{
		\begin{tabular}{c c c c c c c c c}
			\hline\hline \\[-3mm]
			Parameters & Case 1 & Case 2 & Case 3 & Case 4 & Case 5  & Case 6&Case 7\\ \hline
			$k_p$  & 0.0986 & 0.0986 &0.0986&0.0986 & 0.0736 &0.4177&0.5671\\
			$k_q$ & 0.0048 &  0.0048 & 0.0048 &0.0048 & 0.0788&0.081&0.1413\\ 
			$k_{11}$  & 3.1326 & 0.7832 &1.0027&0.2507 & 1.1707&4.1083&5.4297\\
			$k_{12}$ & -0.0104 &  -0.0026 & -0.0033&-0.0008 &-0.0614&-0.0182&-0.0247\\ 
			$k_{13}$ & 0.0155 & 0.0102 &0.0223&0.0119&0.0217&0.0124&0.0082\\
			$k_{21}$ & 0.037 & 0.0422 &0.0417&0.0434&0.7435&0.0624&0.0603\\
			$k_{22}$ & 13.2493 & 13.2493& 13.2493&13.2493&15.0674&17.7106&18.1712\\
			$k_{23}$ & 0.0168 & 0.0168 &0.0167&0.0168&-0.2254&0.0222&0.0228\\[1.4ex]
			\hline\hline
		\end{tabular}
	}
\end{table*}

\begin{figure*}[!t]\centering
	\includegraphics[width=0.9\textwidth]{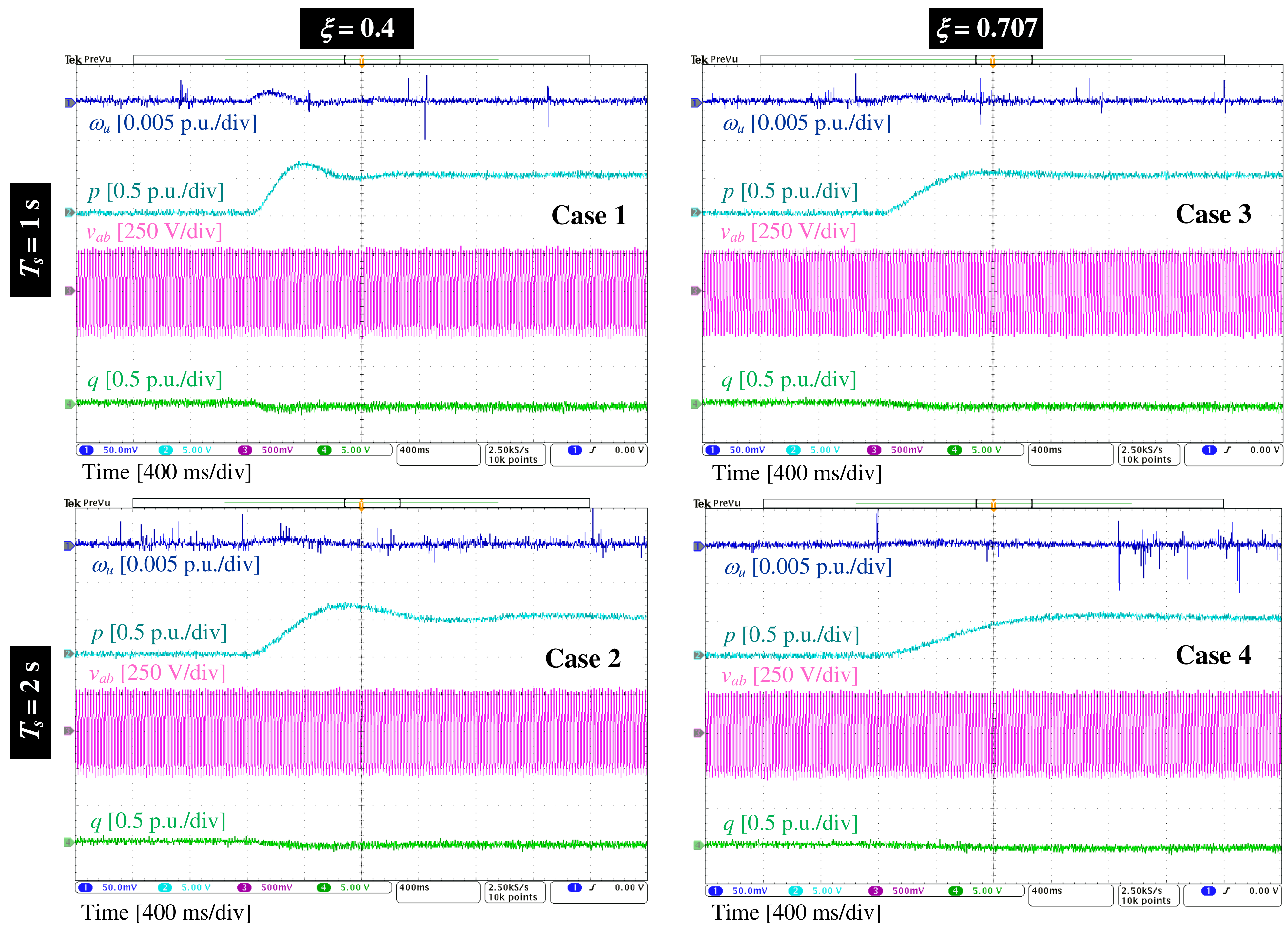}
	\caption{Experimental comparisons with different predefined time-domain performance.}\label{experiment_time_domain}
\vspace{-10pt}
\end{figure*}

Fig. \ref{experiment_time_domain} presents the experimental comparisons of the studied cases when $P_{set}$ steps from 0.5 p.u. to 1 p.u.. As shown, when choosing a large damping ratio ($\xi=0.707$ in Case 3 and Case 4), the dynamics have smaller $P.O.$ than those with a small damping ratio ($\xi=0.4$ in Case 1 and Case 2). Meanwhile, when choosing a small settling time ($T_s=1$s in Case 1 and Case 3), the systems can reach the steady-state quicker than those with a large settle time ($T_s=2$s in Case 2 and Case 4). Fig. \ref{experiment_time_domain} proves that the proposed full-state back control structure and parameter design method are effective to regulate time-domain performance of the power loops of the grid-forming converter.

\subsection{Test on Complex Line}
In this section, the dynamics of the proposed controller and parameter design method is tested with a complex (Case 5) line, where Case 3 with a inductive line is chosen as a base case for comparison. Therefore, the time-domain performance is set the same as Case 3 and the tested parameters of the line are listed in Table \ref{case_line}. The controller parameters of Case 5 based on the step-by-step design procedure are also listed in Table \ref{designed_parameter}.

\begin{table}[!t]
	\renewcommand{\arraystretch}{1.3}
	\caption{Studied Cases of Different Line Characteristics}
	\centering
	\label{case_line}
	\resizebox{0.8\columnwidth}{!}{
		\begin{tabular}{c c c}
			\hline\hline \\[-3mm]
			Cases & Line Impedance $Z_g$ & Ratio $X_g/R_g$\\ \hline
			3  & $j0.0982$ p.u. & +$\infty$ \\
			5 & $0.075+j0.0785$ p.u.&  1.0472  \\ [1.4ex]
			\hline\hline
		\end{tabular}
	}
\end{table}

\begin{figure}[!t]\centering
	\includegraphics[width=0.8\columnwidth]{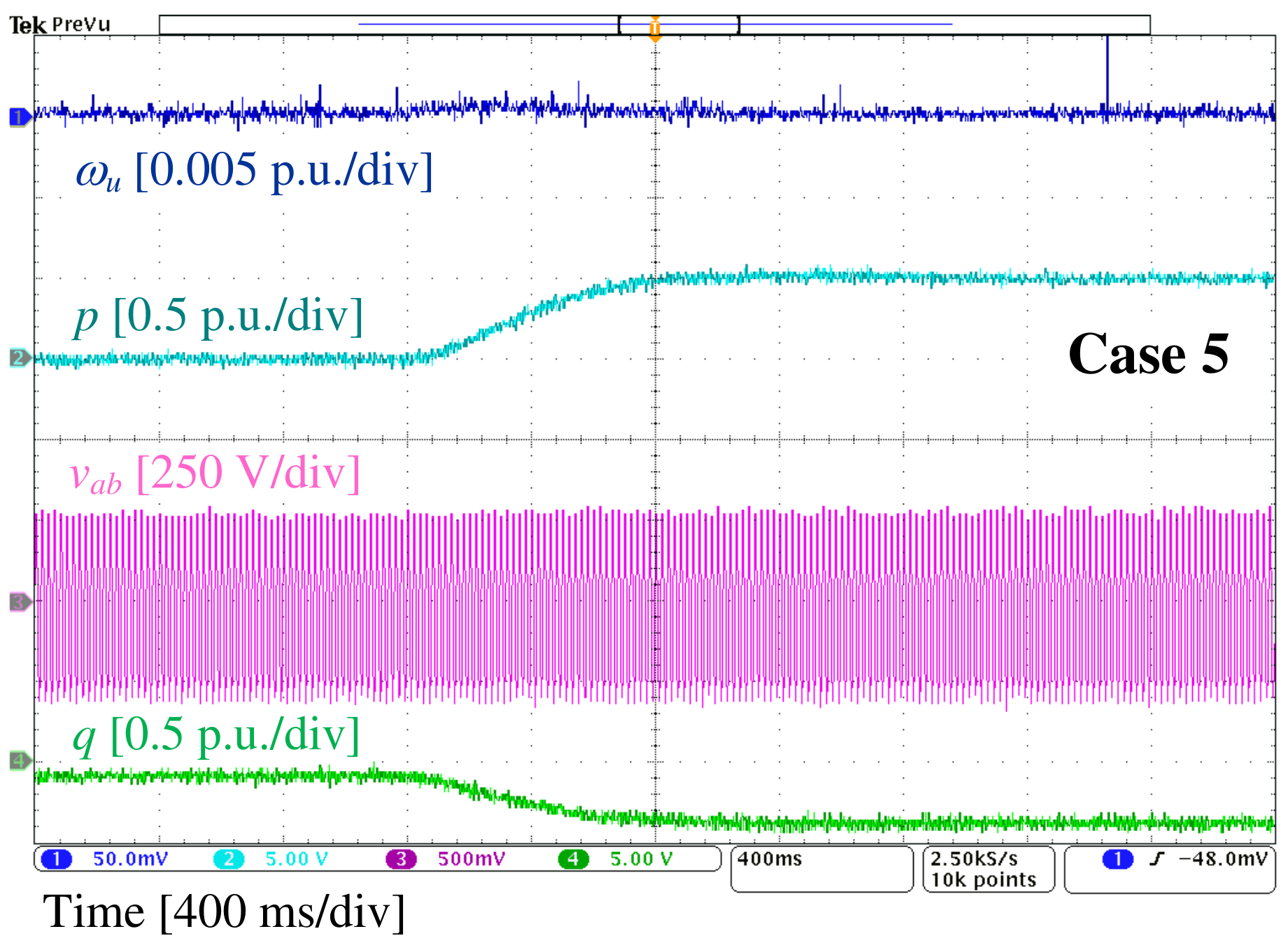}
	\caption{Experimental results with complex line.}\label{experiment_line}
\vspace{-10pt}
\end{figure}

Fig. \ref{experiment_line} shows the experimental waveform. As observed, the dynamics is smooth and in accordance with the predefined time-domain performance, which proves the effectiveness of the proposed method to a complex line.

It should be mentioned that, for a resistive line, the proposed method is completely effective by just using the $p-V$ and $q-f$ droops instead, which will not be presented here.

\subsection{Test on Different Grid Strengths}
\begin{figure}[!t]\centering
	\includegraphics[width=0.8\columnwidth]{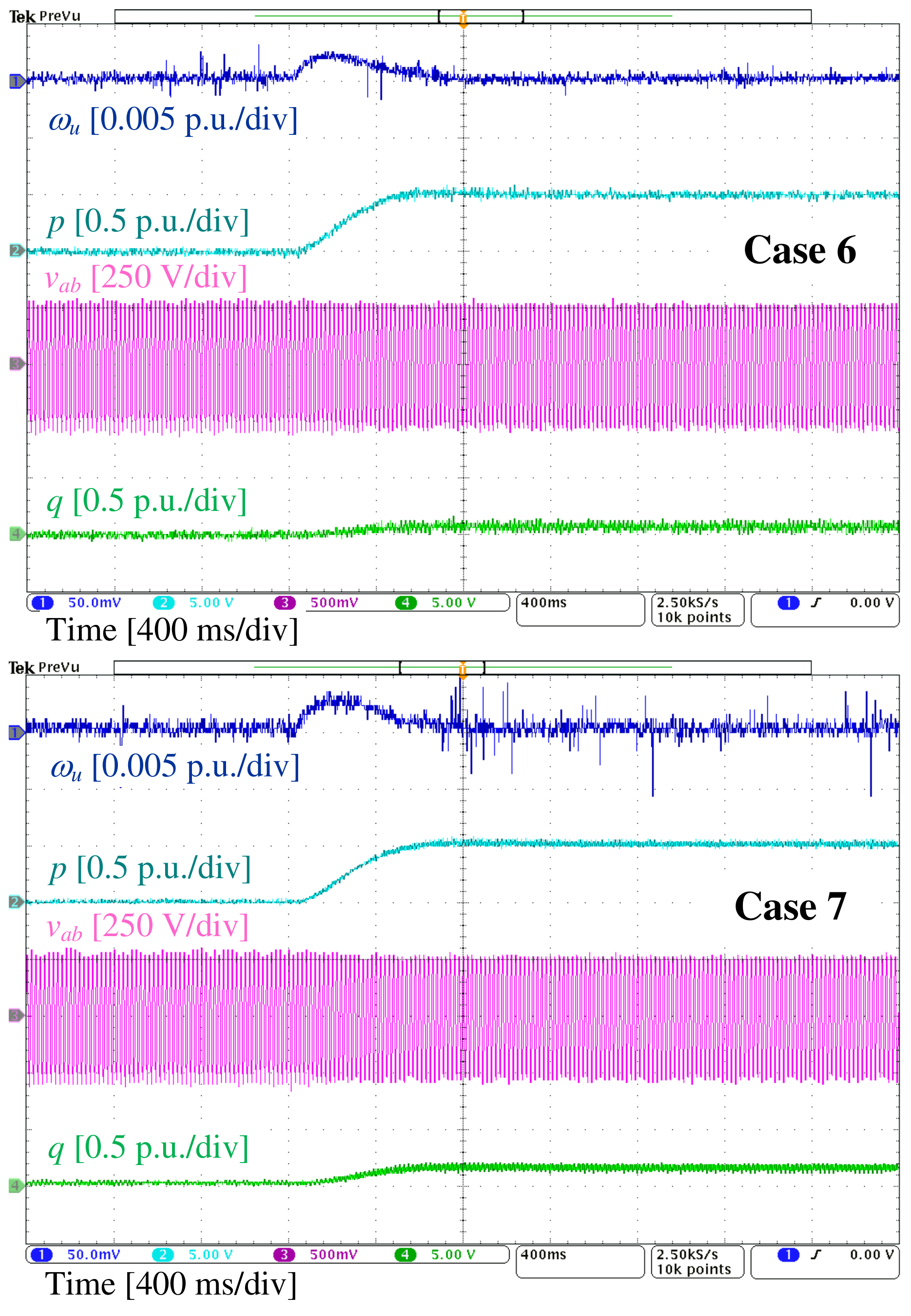}
	\caption{Experimental comparisons with different SCRs.}\label{experiment_scr}
\vspace{-10pt}
\end{figure}

In this section, the dynamics of the proposed controller and parameter design method is tested with weak (Case 6) and very weak (Case 7) grids, respectively, where Case 3 (strong grid) is still chosen as a base case for comparison. The tested parameters of the SCR are listed in Table \ref{case_scr}. The controller parameters of Case 6 and 7 based on the step-by-step design procedure are listed in Table \ref{designed_parameter}, and the experimental comparisons are presented in Fig. \ref{experiment_scr}. As shown, the proposed method has good robustness to the grid strengths.

\begin{table}[!t]
	\renewcommand{\arraystretch}{1.3}
	\caption{Studied Cases with Different SCRs}
	\centering
	\label{case_scr}
	\resizebox{0.7\columnwidth}{!}{
		\begin{tabular}{c c c}
			\hline\hline \\[-3mm]
			Cases & Line Impedance $Z_g$ & SCR\\ \hline
			3  & $j0.0982$ p.u. & 10.1859 \\
			6 & $j0.3927$ p.u.&  2.5465  \\ 
			7 & $j0.5105$ p.u.& 1.9588  \\[1.4ex]
			\hline\hline
		\end{tabular}
	}
\end{table}

It is worth mentioning that, the designed results of Table \ref{designed_parameter} are based on the nameplate values, which may deviate from the actual ones. For example, the actual value of an 1 mH inductor (without power) is about 0.9 mH according to the measurement in the lab, which implies a 10\% error. Meanwhile, the equivalent resistance is neglected as well. Therefore, the presented waveform also verifies that the proposed method has a certain ability against variations of the parameter.

\section{Conclusion}

This paper investigates the grid-forming converter power loops from the perspective of MIMO systems. An error-based regulator model is built by considering the potential coupling of power loops and steady-state droops. Thereafter, the controllability is studied, which reveals the ability of the inputs to control the states. A full-state feedback control strucuture and a step-by-step pole placement-based parameter design method are proposed, which can arbitrarily locate the eigenvalues of the system. The experimental results verify that the work of this paper is effective to cope with different time-domain performance, line characteristics, and grid strengths.

\bibliographystyle{Bibliography/IEEEtran}
\bibliography{Bibliography/IEEEabrv,Bibliography/IEEETIE}\

\end{document}